\documentclass[12pt,preprint]{aastex}
\renewcommand{\cite}{\citealp}

\begin{document}

\title{Metal abundances of RR Lyrae stars in the metal rich globular 
cluster
NGC~6441\altaffilmark{1}}

\author{
Gisella Clementini,\altaffilmark{2}
Raffaele G. Gratton,\altaffilmark{3}
Angela Bragaglia,\altaffilmark{2}
Vincenzo Ripepi,\altaffilmark{4}
Aldo F. Martinez Fiorenzano\altaffilmark{3}$^{,}$\altaffilmark{5} 
Enrico V. Held,\altaffilmark{3}
Eugenio Carretta\altaffilmark{2}
}

\altaffiltext{1}{Based on data collected at the Very Large Telescope of
the European Southern Observatory, Paranal, Chile, program number 71.B-0621.
}

\altaffiltext{2}{INAF, Osservatorio Astronomico di
Bologna, via Ranzani 1, I-40127 Bologna, Italy;
(gisella.clementini, angela.bragaglia, eugenio.carretta)@bo.astro.it}

\altaffiltext{3}{INAF, Osservatorio Astronomico di
Padova, vicolo dell'Osservatorio 5, I-35122 Padova, Italy;
(gratton, held, fiorenzano)@pd.astro.it}

\altaffiltext{4}{INAF, Osservatorio Astronomico di Capodimonte, 
via Moiarello 16, I-80131 Napoli, Italy,
ripepi@na.astro.it}

\altaffiltext{5}{Dipartimento di Astronomia, Universit\`a di
Padova, vicolo dell'Osservatorio 2, I-35122 Padova, Italy}

\begin{abstract}
Low resolution spectra 
have been used to measure individual metal abundances of RR Lyrae stars 
in NGC\,6441, a Galactic globular cluster known to have very unusual
horizontal branch morphology and periods of the RR Lyrae stars for its high
metallicity. 
We find an average metal abundance of [Fe/H]=$-0.69 \pm 0.06$ (r.m.s.=0.33 dex)
and [Fe/H]=$-0.41 \pm 0.06$ (r.m.s.=0.36 dex) on Zinn \& West and Carretta \&
Gratton metallicity scales, respectively, consistent with the 
cluster metal abundance derived by Armandroff \& Zinn. 
Most of the metallicities were extrapolated from calibration relations
defined for [Fe/H]$\leq -1$; however, they are clearly high
and contrast with the rather long periods of the NGC\,6441 
variables, thus confirming that the cluster does not fit in the general 
Oosterhoff classification scheme.
The r.m.s. scatter of the average is larger than 
observational errors (0.15-0.16 dex) possibly indicating 
some spread in metallicity.
However, even the metal poor variables, if confirmed to be cluster
members, are still more metal rich
than those commonly found in
the Oosterhoff type II globular clusters.
\\

\end{abstract}


\section{Introduction}
NGC\,6441, as well as its twin NGC\,6388, are metal rich, massive globular clusters 
([Fe/H]=$-0.53 \pm 0.11$, and $-0.60 \pm 0.15$, respectively: 
\citealt{armandroff88}) with very unusual
horizontal branches (HBs) extending from stubby red, as expected for their high
metallicities, to extremely blue, and with the red HBs sloping upward as one 
moves blueward in the
$V,B-V$ color magnitude diagram (\citealt{rich97}; 
\citealt{pritzl01,pritzl02,pritzl03}).
Given their 
high metallicities, we would expect them to  
have very few RR Lyrae stars 
with the short periods typical of the
Oosterhoff type I systems 
\citep{oosterhoff39}. Rather unexpectedly,  
large numbers of RR Lyrae stars with unusually long periods,
even longer than those commonly observed in the Oosterhoff type II systems,  
 have been discovered in both clusters
(\citealt{layden99,pritzl01,pritzl02,pritzl03}).
Indeed, NGC\,6441 and NGC\,6388 seem to violate the trend of decreasing period with 
increasing metallicity followed by the Galactic globular clusters (GCs), and 
stand apart in the mean period vs. [Fe/H] diagram (\citealt{pritzl01}). 
The clusters have been suggested to be a further and 
extreme manifestation of the so called second-parameter effect, meaning that 
metallicity is not the only factor governing the morphology of the HB, but 
other parameters such as age, helium or CNO abundances, core rotation, 
or dynamical interactions are at work. 
Some of these possibilities (e.g. a high helium abundance, 
higher interaction rates, etc.)  
are not supported by observations for NGC\,6441 and NGC\,6388
(\citealt{rich97,layden99,raimondo02}; Moehler, Sweigart, \& Catelan 1999). 
Thus, we still 
lack a satisfactory explanation for the cluster's peculiar HBs and for
the unusual properties of their RR Lyrae stars.

A metallicity spread,     
as first argued by \citet{piotto97} from the intrinsic spread in color of
the
Red Giant Branch, 
with the RR Lyrae and the blue HB stars being at the metal-poor tail 
([Fe/H]$\lesssim -1.6$) of the cluster's metallicity distributions,  
could in principle explain their anomalous HBs
(\citealt{sweigart02,ree02}). 
In this scenario NGC\,6441 and NGC\,6388, the two most massive 
Galactic GCs after  $\omega$ Cen and M\,54, 
have been suggested to be
the relics of
disrupted dwarf galaxies \citep{ree02},  
similarly 
to $\omega$ Cen which has   
a metallicity spread, and to M\,54 which is considered the
nucleus of the Sagittarius galaxy \citep{layden00}.
However, even the metallicity spread does not completely  
explain the
unusual nature of NGC\,6441 and NGC\,6388 RR Lyrae stars (see discussions in 
\citealt{pritzl01,pritzl02}).
On the other hand, if the RR Lyrae in NGC\,6388 and NGC\,6441 are metal rich,
 they would form a new, distinct subclass of 
long-period, metal rich RR Lyrae 
stars \citep{layden99}, that has no counterpart among the field and cluster RR Lyrae stars known
so far,
except perhaps V9 in 47 Tuc (Carney, Storm, \& Williams 1993).

No direct measure of the metal abundance of the RR Lyrae stars 
in either  cluster existed so far.
\citet{pritzl01,pritzl02}  derived metallicities for some of the {\it ab-}type 
RR Lyrae stars in the two clusters using the parameters of the Fourier 
decomposition of the
light curve and the Jurcsik-Kov\'acs method \citep{jurcsik96,kovacs97}.
They estimated average metal abundances of [Fe/H]=$-0.99$ and $-1.21$
for NGC~6441 and NGC\,6388, respectively, that according to
\citet{jurcsik95} correspond to $-1.3$ and $-1.4$ on the 
Zinn \& West (1984, hereafter ZW84)
metallicity scale. These metallicities are much lower than the 
cluster metal abundances derived by 
\citet{armandroff88}, but are 
close to the metallicity of 
\citet{sweigart02} models which yield a best-fitting model of NGC\,6441
HB for a [Fe/H]$\sim -1.4$ and an $\alpha$ enhancement of
[$\alpha$/Fe]=+0.3.
However, given the uncertainty of the
Jurcsik-Kov\'acs method 
(see discussions in
\citealt{difabrizio05}; \citealt{gratton04}, hereafter G04; and \citealt{clementini05},
hereafter C05), 
and the unusual nature of the NGC~6441 and NGC\,6388 RR Lyrae stars, there is some question as
to the validity of these metallicity determinations.
 
In this Letter we present the first direct measure of 
metallicity 
for RR Lyrae stars in NGC~6441 through spectroscopy, 
and provide
the first
quantitative assessment that the cluster variables are indeed metal rich,
with a few outliers possibly suggesting some spread in metallicity. 
\section{The data}

Spectra of 12 RR Lyrae stars in NGC~6441 were obtained in  
July 2003, 
in the course of a spectroscopic survey of RR Lyrae stars in the Sculptor
dwarf spheroidal galaxy (C05). 
Observations were performed using the 
FORS2 spectrograph 
at the
ESO Very Large Telescope 
(VLT, Paranal, Chile). 
We observed two 
$6.8\arcmin \times 6.8 \arcmin$ 
FORS2 subfields of NGC~6441 centered at $\alpha_{2000}$=$17^h 50^m 19.9^s$,
$\delta_{2000}$=$-37^\circ 05^\prime 21.9^{\prime\prime}$; 
$\alpha_{2000}$=$17^h 50^m 24.5^s$, $\delta_{2000}$=$-37^\circ 00^\prime 
06.3^{\prime\prime}$, 
and comprising 5 and 7 RR Lyrae stars, respectively.
Spectra were collected using 
slits 1$^{\prime\prime}$ wide, 
and about 14$^{\prime\prime}$ long to allow for
sky subtraction.
With this configuration, each pixel corresponds to 0.75\AA.
Our  
wavelength range 
contains both the Ca~{\sc ii} K and the hydrogen 
Balmer lines up to H$\beta$. 
Exposure time was of 300 sec, as an optimal
compromise between S/N and time resolution of the light curve.
Details on the observations and data 
reduction procedures can be found 
in C05.

\begin{figure}\plotone{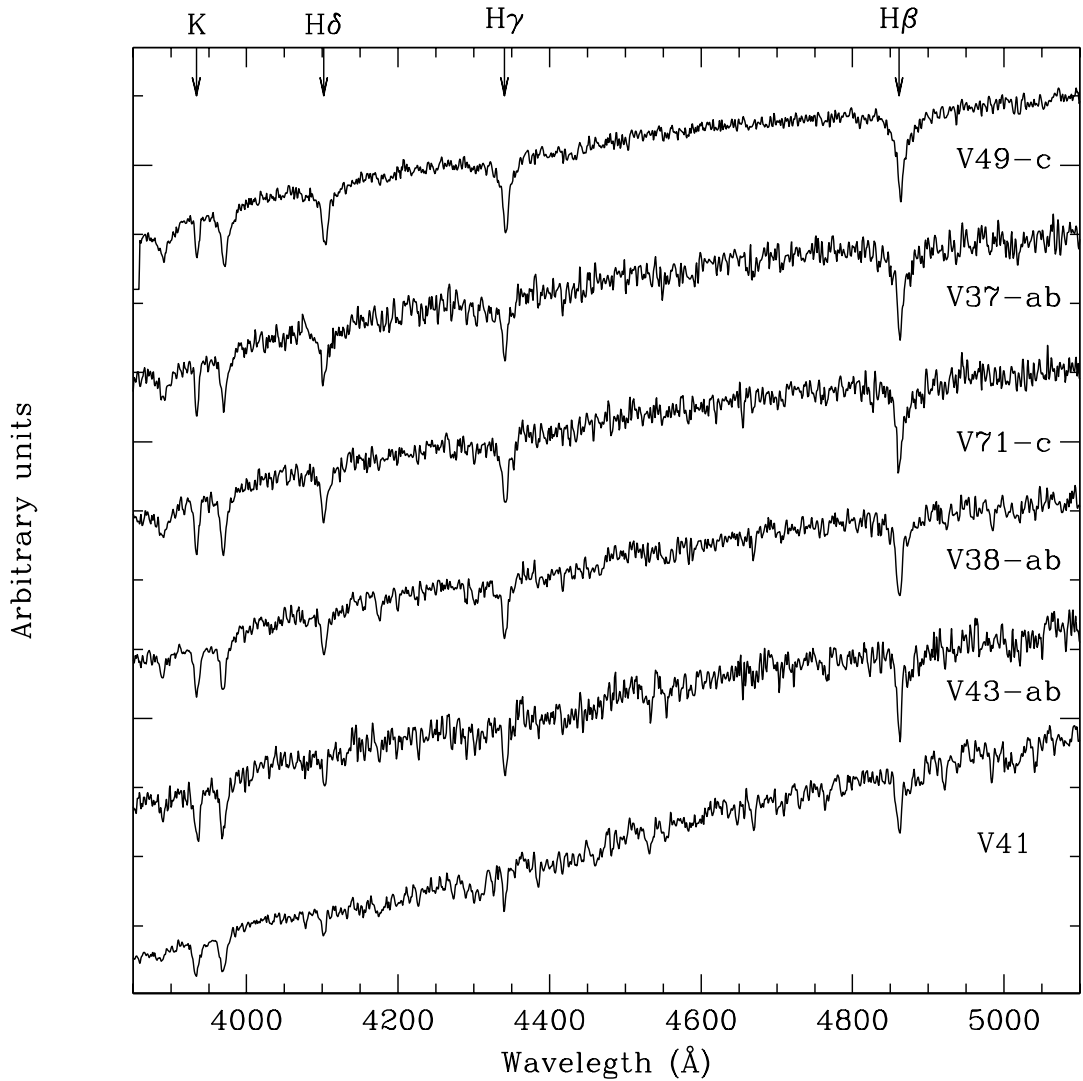} 
\caption{Examples of spectra of 
variable stars  
in NGC\,6441 obtained using FORS2. 
The spectra have been offset for clarity, and the main spectral lines
are indicated. They all
correspond to variable stars having 
$\langle H\rangle < $0.25. 
V41 is very likely the blend
of an RR Lyrae star with a cooler companion.}
\label{f:fig1}
\end{figure}

Time series photometry for all the target stars and classification 
in types has been published by 
 \citet{layden99} and \citet{pritzl01}.
Accordingly, our sample includes 5 {\it ab-} and 7 {\it c-}type RR Lyrae stars;
their identification is provided in Table~\ref{t:NGC6441}.
Figure~\ref{f:fig1} shows examples of spectra  
for some of the variable stars   
in the cluster.

We estimated radial velocities from our spectra; they are given in Column 
12 of Table~\ref{t:NGC6441}.
According to C05 typical errors of these radial velocity
determinations are of about 15 km s$^{-1}$.
Our radial velocity estimates do not exclude the cluster 
membership for any of the RR Lyrae stars we have analyzed.
The 12 stars have $\langle v_{r} \rangle =-$1 km s$^{-1}$  (r.m.s. = 12 
km s$^{-1}$, and zero point error of $\pm$7 km s$^{-1}$).
Our average radial velocity differs somewhat from the value of
+16.4 km s$^{-1}$ (\citealt{harris96}, on line catalogue, available at 
http://www.physics.mcmaster.ca/Globular.html).
A reason for part of this discrepancy is that 
our mean value includes the phase-dependent 
contributions due to 
the star pulsations. 
Further residual differences may be due 
to systematic offsets possibly caused by offcentering of the
cluster variables on the slits.

\section[]{Measure of the metal abundances} 

Metal abundances for the RR Lyrae stars in NGC~6441 
 were derived using a modified (and improved) 
version of the $\Delta$S method \citep{preston59}.
Our technique is fully described in G04,
and is based on
 the definition of Hydrogen and Ca~{\sc ii} spectral indices, 
 $\langle H\rangle$ and $K$, for each variable star by directly integrating 
 the instrumental fluxes in spectral
 bands centered on the  H$\delta$, H$\gamma$, H$\beta$, and 
 Ca~{\sc ii} K lines.
 These spectral indices are then used to measure metallicities by comparison 
  to the same
quantities for variable stars in a number of globular clusters of 
known metal abundance. 
 A summary of the method and an update of the 
 metallicity calibration procedure 
can be found in C05.
An advantage of
our technique is that we do not need to know the phases of our spectra. On the
other hand, the accuracy of our [Fe/H] may be a function of phase, as 
represented
by the strength of the H lines. According to figure 12 of G04, most accurate 
metallicity determinations are obtained for values of $\langle H\rangle <$ 0.20, and
0.25 for {\it ab-} and {\it c-}type RR Lyrae stars, respectively. 
Outside these ranges metal abundances determinations may be more uncertain, 
depending on the actual value of 
$\langle H\rangle $. However, metallicities of individual stars
in the calibrating clusters (see Tables 3, 4 in G04, and Tables 5, 6, and 7
 in C05) show this effect to be small, if present. 
This point is particularly relevant for the NGC\,6441 RR Lyrae stars
for which, based on the available photometric data \citep{layden99, pritzl01}
it may be difficult to reliably define the pulsation phase at the epochs 
of the observations.
Line indices measured for the RR Lyrae stars in NGC~6441 
are provided in Columns
from 4 to 8 of Table~\ref{t:NGC6441}. The stellar K values 
were not corrected for the interstellar K lines contribution
 since we estimated it to be 
 much less than 0.1 dex in [Fe/H].
In fact, given the small absolute value of the cluster radial velocity, 
interstellar
lines are expected to lie in the core of the stellar K-lines where there is
almost no flux, and
then can subtract only a negligible fraction of flux\footnote{We checked 
this point on the spectra of a few NGC\,6441 giants we recently observed
with UVES@VLT. We estimated that, according to the definition of
the spectral indices used in the present study, the interstellar K 
line contribution 
may cause a systematic offset $<$ 0.02 dex in the metal abundance of 
the NGC\,6441 variable stars.}. 
This would not be the 
case for halo stars where, due to the high velocities, the interstellar absorptions
occur outside the line cores.

\begin{figure}\plotone{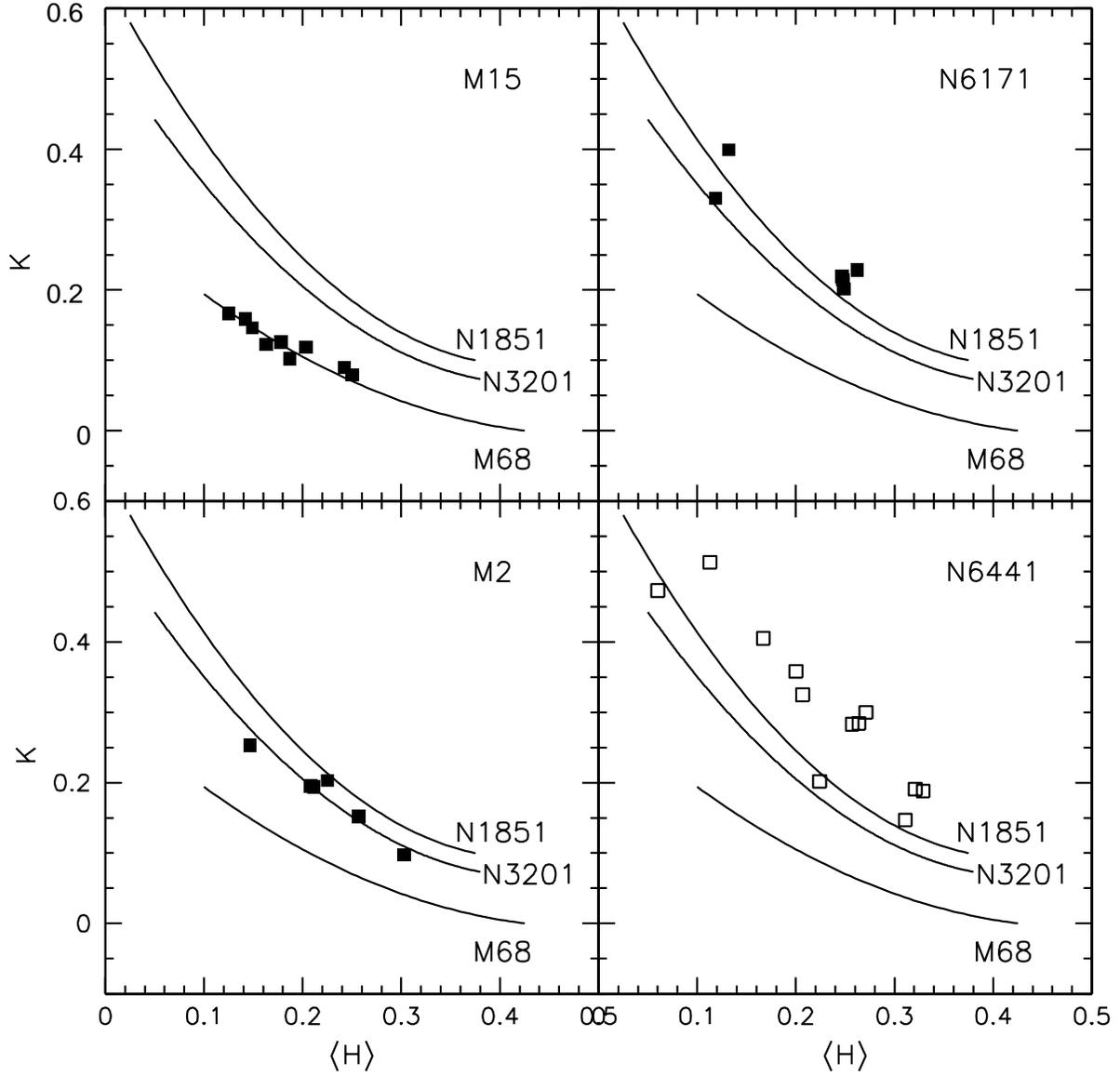} 
\caption{Correlations between $K$ and $<H>$ spectral indices for the
calibrating clusters M\,15, M\,2, and NGC\,6171 (from C05,
solid squares) and  M\,68, NGC\,3201, and NGC\,1851 (from G04,
 solid lines). 
For a definition of the spectral indices, see G04.}
\label{f:fig2}
\end{figure}

The calibration of the line indices of the variable stars 
in terms of metal abundances [Fe/H] was obtained using RR Lyrae stars 
observed 
in the clusters M\,15, M\,2 and NGC\,6171 (C05), and
M\,68, NGC\,1851 and NGC\,3201 (G04).
For all these clusters, precise metal abundances are available on both 
the ZW84 and the Carretta \& Gratton (1997, hereafter CG97) 
metallicity scales. 
Figure~\ref{f:fig2} shows 
the correlation between $K$ and $\langle H\rangle$ indices 
for the calibrating clusters and, in the bottom-right panel, the position 
on the $K$ vs $\langle H\rangle$ plane of the NGC\,6441 variable stars. 
The figure shows 
that there are 
three objects lying near 
the ridge line of NGC\,1851 at an average metallicity
of [Fe/H]$\simeq -1.3$ on the ZW84 scale, while all
 the remaining stars define a tight correlation at higher 
metallicity. However, one of the deviating objects, star
V41, is probably blended with a
cooler companion, as suggested by its spectrum (see Fig. 1). Thus its 
metallicity is uncertain and we will drop it from any further consideration. 
 V49 was observed in the safe range where metal abundance
determinations are most reliable.
The spectrum of this star is shown in Fig.~\ref{f:fig1}, along with spectra
of all the variables having $\langle H\rangle < $0.25, and clearly shows that 
the star has a shallower $K$ line.
Finally, V74 was observed at $\langle H\rangle$=0.311, thus  
its metallicity may be slightly more uncertain. 

\begin{figure}\plotone{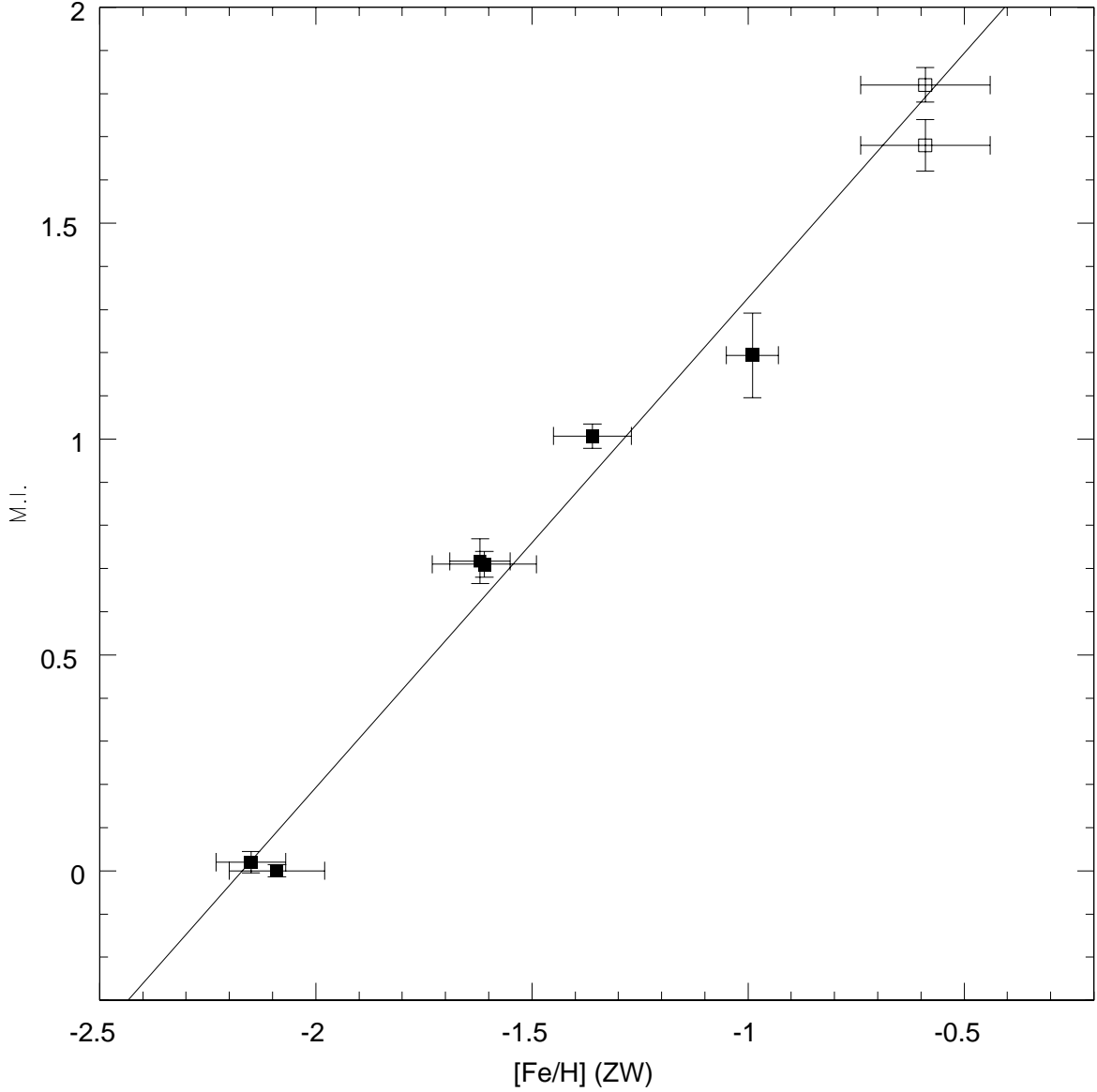}
\caption{Calibration of the metallicity index (M.I.) on the ZW84 metallicity
scale. Open squares represent NGC\,6441, which is plotted 
according to its metallicity from ZW84 ([Fe/H]=$-0.59$) and the M.I. values
obtained both averaging all the 11 variable stars (lower open square), and 
discarding the two metal poor objects V49 and V74 (upper open square). 
In both cases the cluster lies very close to the calibration line, which is 
based on the other clusters only, and perfectly falls 
on it if V49 and V74 are disregarded.
}
\label{f:cal02}
\end{figure}

Metallicity indices, $M.I.$'s, for the program stars were 
derived from their $K$ and $\langle H\rangle$ values using 
equations (3), (4), and (5) of G04. 
They 
are listed
in Column 9 of Table~\ref{t:NGC6441}.
Metal abundances [Fe/H] were then deduced from the $M.I.$ values 
using the metallicity calibrations defined by 
C05. These are described by linear regressions,
namely equations (4) and (5) of 
C05, between the 
average $\langle M.I.\rangle$ values of RR Lyrae stars 
in the calibrating clusters M\,15, M\,2, NGC\,6171, M\,68, NGC\,1851 and NGC\,3201
and the cluster's metal abundances on the ZW84 and 
 CG97 metallicity scales,
respectively.
In Figure~\ref{f:cal02} we show the calibration relation of the
metallicity indices in the ZW84 metallicity
scale.   
We note that the calibrating clusters have metallicities that do not extend 
higher than [Fe/H]$\sim -1$, thus the metal abundance we derive for NGC\,6441
is an extrapolation of the calibration equations. However, the cluster is
found to fall 
well on the extrapolation to higher metallicities of 
the linear relation defined by the calibrating clusters. This results is in 
agreement with
the original Preston's $\Delta$S method, which appears to have a linear 
calibration up to
about solar metallicity.  
Individual metal abundances  
derived with this procedure are given in Columns 10 and 11 of 
Table~\ref{t:NGC6441}. According to C05
we attach internal errors of 0.15 and 0.16 dex to individual abundance
determinations in ZW84 and CG97 scale, respectively.
However, to take into account any additional uncertainty which might 
affect the spectra corresponding to $\langle H\rangle $ values outside the 
safest range, in the following we will 
divide the stars in three groups corresponding to $\langle H\rangle \leq $0.224,
0.257 $ \leq \langle H\rangle \leq $ 0.271, and 0.311 $ \leq \langle H\rangle 
\leq $ 0.329, and all averages will be computed giving different weights to the
3 groups of stars (i.e. 1, 0.75 and 0.5). 

For comparison, in the last column of Table~\ref{t:NGC6441} we report 
the metal abundances derived by \citet{pritzl01} from the Fourier 
decomposition of the light curve for 4 of the {\it ab-}type RR Lyrae stars
analyzed here. 
The average of \citet{pritzl01} values 
for the 4 stars is [Fe/H]=$-1.02 ~~(r.m.s.=0.15)$,  in 
\citet{jurcsik95} metallicity scale. According to equation
(4) in \citet{jurcsik95} this corresponds to [Fe/H]=$-1.33$, on the ZW84 scale. 
The average of our metal abundances for these 4 stars 
is [Fe/H]=$-0.67 ~~(r.m.s.=0.11)$, on the ZW84 scale, i.e. 
much more metal rich.
This result seems to suggest that the
Jurcsik-Kov\'acs method may not be reliable when applied 
to the anomalous RR Lyrae stars in NGC\,6441.

Adopting the averaging scheme described above, 
the mean metal abundance of our RR Lyrae sample in NGC\,6441 is 
[Fe/H]=$-0.69 \pm 0.06$  (r.m.s.=0.33 dex, 11 stars), and $-0.41 \pm 0.06$  
(r.m.s.=0.36 dex) in ZW84 and CG97 scale, respectively. 
The scatter
of the average is larger than expected from measurement errors 
alone ($<$0.2 dex) thus 
suggesting that, if the 11 variables are all cluster members,
there is some spread in 
metallicity 
with two more metal poor objects at an average metal abundance around
[Fe/H]$\sim -1.3$, and the remaining 9 more metal rich stars 
around [Fe/H]$\sim -0.6$ dex (on the ZW84 scale). If the two metal
poor stars are disregarded, the mean metal abundances become
[Fe/H]=$-0.57\pm 0.04$  (r.m.s.=0.19 dex, 9 stars), and $-0.28 \pm 0.04$  
(r.m.s.=0.21 dex).
The former value is in excellent agreement with both ZW84 and 
\citet{armandroff88} metallicity estimates for the cluster.

\section[]{Summary and conclusions}
Metal abundances from low resolution spectroscopy obtained with FORS2 at the
ESO VLT have revealed that the NGC\,6441 RR Lyrae stars are metal rich, with an
average metal abundance of [Fe/H]=$-0.69 \pm 0.06$, on ZW84 scale.
The spectroscopic analysis also reveals 
that there are two variables (out of eleven) having metallicities  
around [Fe/H]$\sim -1.3$.
However, the metal poor stars, if confirmed cluster members, are a
minor component of our sample.
Even allowing for 
measurement errors, we do not find in this cluster RR Lyrae stars as metal poor 
as the variables 
commonly found in the Galactic Oosterhoff type II GCs,
as instead one would expect from the extraordinarily long periods and the position near
the Oosterhoff II line of the NGC\,6441 variables in the period-amplitude diagram
(see figure~6 of \citealt{pritzl03}).
Clearly, metal abundances and memberships for a larger number of variable 
stars are needed to
better assess the metallicity distribution of the NGC\,6441 stars and 
the relevance of the metal poor component, if any. Nevertheless,  
 the existence of extremely long period RR Lyrae stars with extraordinarily 
 high metal abundances, as some of the variables in our sample,   
demonstrates that the NGC\,6441 variable stars are 
different from the RR Lyrae stars 
known so far both in the Milky Way GCs and in the field, and confirms 
that this cluster does  
not conform to the Oosterhoff dichotomy described by the other Galactic 
GCs.  

Which mechanism may be able to produce such metal 
rich variables
with pulsation characteristics similar to the Oosterhoff type II ones
remains unexplained. 
For instance, \citet{pritzl02} show that it is difficult to  model NGC\,6441 as an
Oosterhoff II system under the hypothesis that its variables are evolved from a
position on the blue zero-age HB, as a result of the small number of 
progenitors on the blue HB. As suggested by
\citet{layden99} and \citet{pritzl01},  
the theoretical 
reproduction of the observed light curves with pulsation models may 
shed some light on the physical properties
responsible for the anomalous properties of the NGC\,6441 RR Lyrae stars.
Such modeling for the variables analyzed in the present paper 
is currently under way (Clementini \& Marconi 2005, in
preparation). 
%
%


\acknowledgments 
This research was funded 
by MIUR, under the scientific project:
2003029437, ``Continuity and Discontinuity in the 
Milky Way Formation'' (P.I.: Raffaele Gratton).


\clearpage

\begin{deluxetable}{lccccccccccrc}
\rotate
\tablewidth{0pc}
\tablecaption{Line indices and metal abundances of RR Lyrae stars in the globular cluster NGC\,6441\label{t:NGC6441}}
\tabletypesize{\tiny}
\tablehead{
\colhead{~~{\rm Star}} & 
\colhead{{\rm HJD}} & 
\colhead{{\rm Type}} & 
\colhead{{\rm K}} & 
\colhead{{\rm H$_{\delta}$}} & 
\colhead{{\rm H$_{\gamma}$}} & 
\colhead{{\rm H$_{\beta}$}} & 
\colhead{$\langle$ {\rm H} $\rangle$}  & 
\colhead{{\rm M.I.}} & 
\colhead{{\rm [Fe/H]}}   &   
\colhead{{\rm [Fe/H]}} & 
\colhead{{\rm V$_{r}$}~~}&
\colhead{{\rm [Fe/H]}} \\
\colhead{~~(a)}& 
\colhead{($-$2400000)} &
\colhead{} & 
\colhead{}   &
\colhead{}  &
\colhead{}  &
\colhead{}  &
\colhead{}  &
\colhead{}  &   
\colhead{{\rm ZW84}}   &  
\colhead{{\rm CG97}} & 
\colhead{{\rm km~ s$^{-1}$}} &
\colhead{{\rm P01}}
}
\startdata
 {\rm ~V37}              & 52850.6112 &{\rm ab}& 0.325 & 0.216 & 0.213 & 0.193 & 0.207 & 1.66 & -0.71 & -0.44 & +18~~&-0.95\\
 {\rm ~V38}              & 52850.6112 &{\rm ab}& 0.405 & 0.157 & 0.155 & 0.191 & 0.167 & 1.69 & -0.68 & -0.41 & -7~~&-1.14\\
 {\rm ~V41\tablenotemark{b}} & 52849.6208 &{\rm ab}& 0.473 & 0.056 & 0.030 & 0.096 & 0.060 & 0.91 & -1.37 & -1.15 & +4~~&\nodata\\
 {\rm ~V43}              & 52850.6112 &{\rm ab}& 0.513 & 0.098 & 0.119 & 0.122 & 0.113 & 1.60 & -0.76 & -0.49 &  -4~~&-1.16\\
 {\rm ~V49\tablenotemark{c}} & 52850.6112 &{\rm c}& 0.202 & 0.253 & 0.220 & 0.199 & 0.224 & 0.90 & -1.37 & -1.15 & +22~~&\nodata\\
 {\rm ~V51}              & 52850.6112 &{\rm ab}& 0.283 & 0.265 & 0.270 & 0.236 & 0.257 & 1.94 & -0.46 & -0.11 &-16~~&-0.84\\
 {\rm ~V69}              & 52849.6208 &{\rm c} & 0.300 & 0.294 & 0.260 & 0.260 & 0.271 & 2.29 & -0.15 &  +0.16 &-10~~&\nodata\\
 {\rm ~V71}              & 52850.6112 &{\rm c} & 0.358 & 0.222 & 0.187 & 0.192 & 0.200 & 1.80 & -0.58 & -0.30 & +10~~&\nodata\\
 {\rm ~V72}              & 52850.6112 &{\rm c} & 0.191 & 0.340 & 0.349 & 0.275 & 0.321 & 1.72 & -0.65 & -0.38 & -6~~&\nodata\\
 {\rm ~V74}              & 52849.6208 &{\rm c} & 0.147 & 0.331 & 0.312 & 0.289 & 0.311 & 1.17 & -1.14 & -0.90 & -7~~&\nodata\\
 {\rm ~V77}              & 52849.6208 &{\rm c} & 0.284 & 0.275 & 0.298 & 0.220 & 0.264 & 2.05 & -0.36 & -0.07 & -12~~&\nodata\\
 {\rm ~V78}              & 52849.6208 &{\rm c} & 0.188 & 0.347 & 0.356 & 0.286 & 0.329 & 1.75 & -0.63 & -0.36 & -12~~&\nodata\\
\enddata
\tablenotetext{a}{Identifiers are from Sawyer-Hogg on line catalogue of variable stars in Galactic
globular clusters published by \citet{clement01}}

\tablenotetext{b}{\citet{layden99} find that this star is $\sim$ 0.7 mag brighter and
$\sim$ 0.3 mag redder than the other RR Lyrae stars and suggest its image is blended
with a red companion. \citet{pritzl01} agree on this hypothesis.
The star is in the cluster central region and from our spectrum it looks cooler 
than a normal RR Lyrae (see Figure~\ref{f:fig1}), as also demonstrated by its rather 
small $\langle H \rangle$ value, thus supporting the blend hypothesis.}

\tablenotetext{c}{This star is a suspected  binary system for \citet{layden99}. However,
\citet{pritzl01} 
find it is a slightly overluminous {\it c-} type RR Lyrae. The star spectrum 
shows typical features
of an RR Lyrae star (see  Figure~\ref{f:fig1}) confirming \citet{pritzl01} classification.}
\end{deluxetable}

\clearpage

\clearpage

\clearpage

\end{document}